\UseRawInputEncoding
\documentclass[pra,twocolumn,shownopacs,amssymb]{revtex4}
\usepackage{graphicx,psfrag,amsmath,amssymb}
    
\begin{document}

\title{Effective-mass tensor of the two-body bound states and the quantum-metric 
tensor of the underlying Bloch states in multiband lattices}

\author{M. Iskin}
\affiliation{Department of Physics, Ko\c{c} University, Rumelifeneri Yolu, 
34450 Sar\i yer, Istanbul, Turkey}

\date{\today}

\begin{abstract}

By considering an onsite attraction between a spin-$\uparrow$ and a spin-$\downarrow$ 
fermion in a multiband tight-binding lattice, here we study the two-body spectrum, and derive 
an exact relation between the inverse of the effective-mass tensor of the lowest bound states 
and the quantum-metric tensor of the underlying Bloch states. 
In addition to the intraband (or the so-called conventional) contribution that depends only 
on the single-particle spectrum and the interband (or the so-called geometric) contribution 
that is controlled by the quantum metric, our generalized relation has an additional 
interband contribution that depends on the so-called band-resolved quantum metric. 
All of our analytical expressions are applicable to those multiband lattices that simultaneously 
exhibit time-reversal symmetry and fulfill the condition on spatially-uniform pairing. 
As a nontrivial illustration we analyze the two-body problem in a Kagome lattice with 
nearest-neighbor hoppings, and show that the exact relation provides a perfect benchmark.

\end{abstract}

\maketitle

\section{Introduction}
\label{sec:intro}

Quantum-metric tensor is defined as the real part of the quantum-geometric tensor 
(whose imaginary part is the Berry curvature), and in condensed-matter physics it 
provides a measure of the so-called quantum distance between the nearby 
Bloch states in momentum space~\cite{provost80, berry89, resta11}. 
Together with the Berry curvature, it constitutes one of the two band-structure 
invariants that have gradually become central objects in a number of fields.
For instance the quantum metric plays a crucial role in the transport properties 
of some multiband superfluids and superconductors: it controls the effective-mass 
tensor of the superfluid carriers through the interband processes, which in return affects 
all of the other superfluid properties that depend on the carrier mass including, e.g., 
the superfluid weight/density, velocity of the low-energy collective modes such as 
the Goldstone mode, BKT transition temperature, 
etc~\cite{peotta15, julku16, torma17a, iskin18, iskin19, wang20, iskin20, wu21}. 
Given these theoretical predictions and many others, there has been a rapid progress 
and growing demand for measuring the quantum metric 
itself~\cite{asteria19, yu20, gianfrate20, tan21, liao21}. 

In connection to the central theme of this paper, we have recently derived a 
Ginzburg-Landau functional for a spin-orbit-coupled Fermi superfluid in 
continuous space (within the 
BCS-BEC meanfield theory and Gaussian fluctuations on top of it), and revealed a 
direct relation between the inverse of the effective-mass tensor of the many-body 
bound states (e.g., Cooper pairs) and the quantum-metric tensor of the helicity 
states~\cite{iskin18}. See also~\cite{iskin19}. 
Then, by assuming a sufficiently weak onsite attraction between 
the particles in a multiband lattice, a parallel relation has been derived for the 
two-body bound states in vacuum by focusing on an isolated flat band that is 
separated from the remaining ones with a finite band gap~\cite{torma18}. 
More recently, by assuming an onsite attraction and time-reversal symmetry, 
we have found an analogous but exact relation for the effective-mass tensor of 
the lowest bound states in generic two-band lattices~\cite{iskin21}. 
In this paper we extend and generalize the latter study to multiband lattices, 
and show that in addition to the intraband (or the so-called conventional) contribution 
that depends only on the single-particle spectrum and the interband (or the so-called 
geometric) contribution that is controlled by the quantum metric, there is an additional 
interband contribution that depends on the so-called band-resolved quantum metric. 
The revelation of the latter contribution is one of our main results in this work. 
All of our analytical expressions are applicable to a certain class of multiband lattices 
that simultaneously exhibit time-reversal symmetry and fulfill the condition on 
spatially-uniform pairing, i.e., when the two-body wave function is uniformly 
delocalized over the sublattices. This is expected to be the case for those Bloch 
Hamiltonians that are invariant under the interchange of their sublattices.
Furthermore we show that our general relation reproduces all of the known results 
in the respective limits, and it is in perfect agreement with the exact solution to the 
two-body problem in a Kagome lattice with nearest-neighbor hoppings.

The rest of the paper is organized as follows. First we study the Kagome model
in Sec.~\ref{sec:kagome}: the single-particle band structure is reviewed in 
Sec.~\ref{sec:bs} and the two-body spectrum is analyzed in Sec.~\ref{sec:tbs}.
Then we relate our numerical findings to the quantum metric in Sec.~\ref{sec:qm},
and discuss the versatility of our results for other lattices in Sec.~\ref{sec:up}.
The paper ends with a brief summary of conclusions and an outlook given in 
Sec.~\ref{sec:conc}.

\section{Kagome Lattice}
\label{sec:kagome}

As a physical motivation for the exact relation between the inverse of the effective-mass 
tensor of the lowest two-body bound states and the quantum-metric tensor of the 
underlying Bloch states, here we want to analyze a nontrivial yet analytically-tractable 
multiband tight-binding lattice that features a flat band in its band structure. 
Given their recent realizations, Kagome~\cite{jo12, nakata12, li18, leung20} and 
Lieb~\cite{diebel16, kajiwara16, ozawa17} lattices are probably the ideal candidates, 
and here we focus on the former model with nearest-neighbor hopping.

\subsection{Band Structure}
\label{sec:bs}

\begin{figure}[htbp]
\includegraphics[scale=0.29]{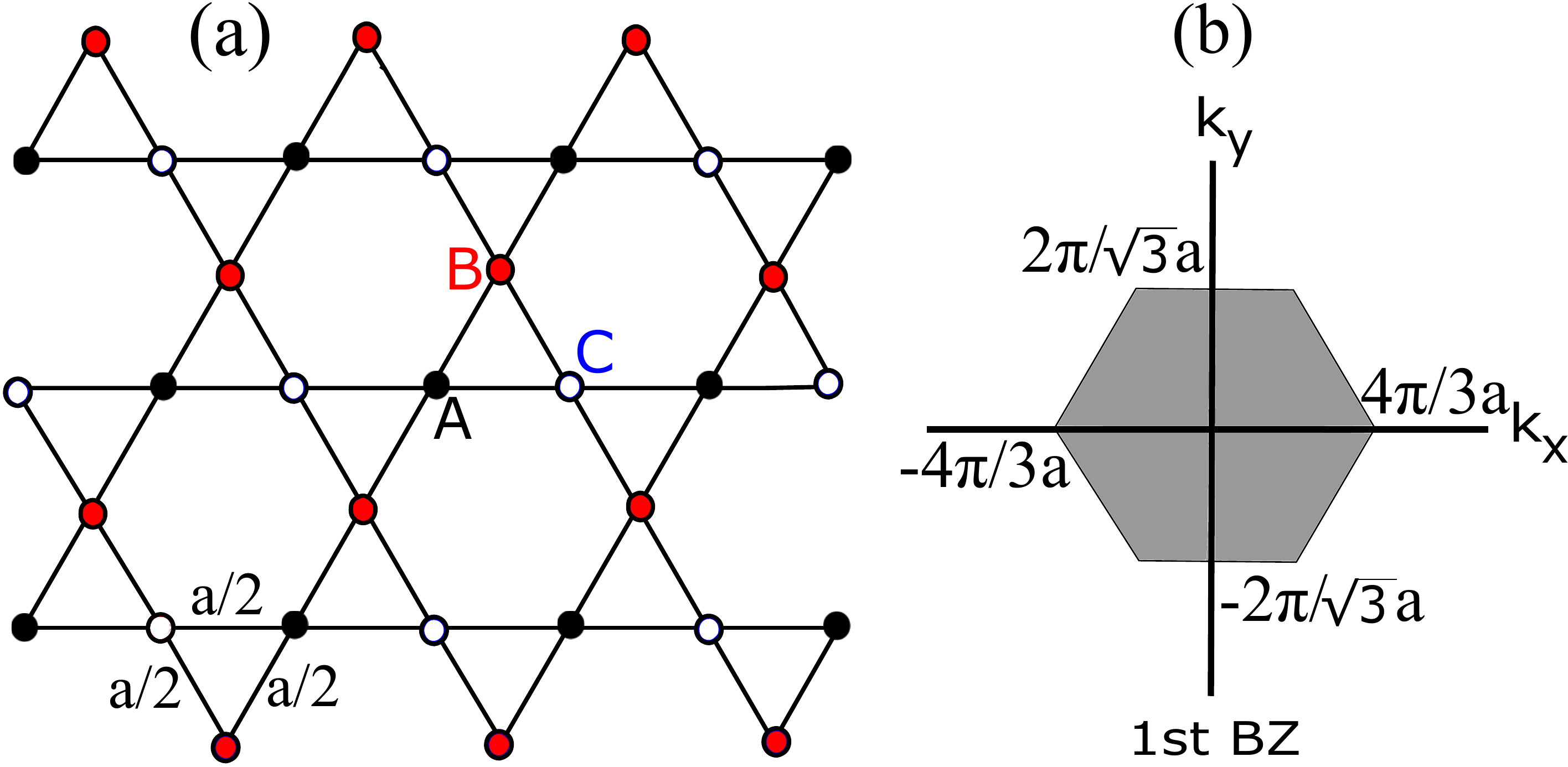}
\centerline{\scalebox{0.43}{\includegraphics{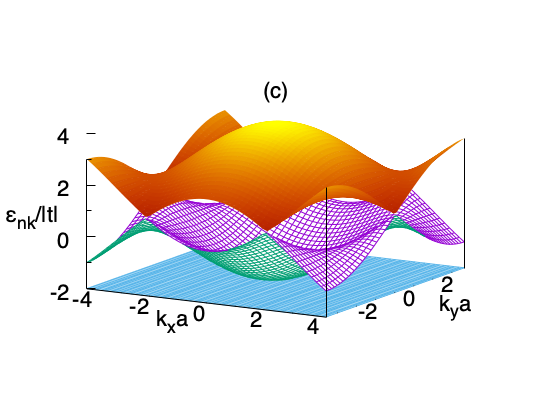}}}
\caption{
\label{fig:bs}
Sketches of (a) the crystal structure in real space and (b) the first Brillouin zone in 
reciprocal space. (c) Band structure with a flat band at the bottom. Note that while
the flat band is in quadratic touch with a dispersive band at the origin, the dispersive 
bands touch each other and form Dirac cones at the six corners of the Brillouin zone.
}
\end{figure}

The Kagome crystal structure with nearest-neighbor bonds is illustrated in 
Fig.~\ref{fig:bs}(a): it forms a triangular lattice with a side length $a$ and has a 
three-point ($N_b = 3$) basis that are located at $\mathbf{r}_A = (0,0)$, 
$\mathbf{r}_B = \frac{a}{4}(1,\sqrt{3})$ and $\mathbf{r}_C = \frac{a}{2}(1,0)$. 
The real-space primitive unit vectors can be chosen as $\mathbf{a}_1 = a(1,0)$ 
and $\mathbf{a}_2 = \frac{a}{2}(1, \sqrt{3})$, and we define 
$\mathbf{a}_3 = \mathbf{a}_1 - \mathbf{a}_2$ for convenience. If the entire lattice 
is constructed with $N_c$ unit cells then the total number of lattice sites is $N = N_b N_c$. 
Accordingly the reciprocal-space primitive unit vectors can be chosen as
$
\mathbf{b}_1 = \frac{2\pi}{\sqrt{3}a} (\sqrt{3},-1),
$
and
$
\mathbf{b}_2 = \frac{4\pi}{\sqrt{3}a} (0,1)
$
and they satisfy
$
\mathbf{a}_i \cdot \mathbf{b}_j = 2\pi\delta_{ij}
$
with $\delta_{ij}$ the Kronecker-delta. The corresponding first Brillouin zone is illustrated 
in Fig.~\ref{fig:bs}(b) whose area is $8\pi^2/(\sqrt{3} a^2)$. This is in such a way that 
the Brillouin zone contains a total of $\sum_\mathbf{k} 1 = N_c$ distinct $\mathbf{k}$-space 
points.

The Hamiltonian for a single spin-$\sigma = \{\uparrow, \downarrow\}$ particle can 
be written as
$
\mathcal{H}_\sigma = \sum_\mathbf{k} \psi_{\mathbf{k} \sigma}^\dagger 
\mathbf{H}_{\mathbf{k} \sigma}
\psi_{\mathbf{k} \sigma},
$
where
$
\psi_{\mathbf{k} \sigma} = ( c_{A \mathbf{k} \sigma}\; c_{B \mathbf{k} \sigma}\; 
c_{C \mathbf{k} \sigma} )^\mathrm{T}
$
is a three-component spinor with $\mathrm{T}$ the transpose operator and $c_{S \mathbf{k} \sigma}$ 
the annihilation operator for a spin-$\sigma$ particle on the sublattice $S$ with a crystal 
momentum $\mathbf{k} = (k_x, k_y)$. In the orbital basis
$
|S \mathbf{k} \sigma \rangle = c_{S \mathbf{k} \sigma}^\dagger | 0 \rangle
$
with $| 0 \rangle$ the vacuum state, the Hamiltonian density can be written as
\begin{align}
\label{eqn:H0k}
\mathbf{H}_{\mathbf{k} \sigma} = -2t
\begin{bmatrix}
0 & \cos\left(\frac{\mathbf{k} \cdot \mathbf{a}_2}{2}\right) 
& \cos\left(\frac{\mathbf{k} \cdot \mathbf{a}_1}{2}\right) \\
\cos\left(\frac{\mathbf{k} \cdot \mathbf{a}_2}{2}\right) & 0 
& \cos\left(\frac{\mathbf{k} \cdot \mathbf{a}_3}{2}\right) \\
\cos\left(\frac{\mathbf{k} \cdot \mathbf{a}_1}{2}\right) 
& \cos\left(\frac{\mathbf{k} \cdot \mathbf{a}_3}{2}\right) & 0
\end{bmatrix},
\end{align}
where $t$ is the hopping element between the nearest-neighbor lattice sites. 
The single-particle spectrum $\varepsilon_{n \mathbf{k} \sigma}$ is given by the 
eigenvalues of $\mathbf{H}_{\mathbf{k} \sigma}$, leading 
to~\cite{barreteau17, mizoguchi19}
\begin{align}
\label{eqn:e1k}
\varepsilon_{1\mathbf{k} \sigma} &= 2t, \\
\label{eqn:e2k}
\varepsilon_{2\mathbf{k} \sigma} &= -t - |t| \sqrt{2 \Lambda_\mathbf{k} + 3}, \\
\label{eqn:e3k}
\varepsilon_{3\mathbf{k} \sigma} &= -t + |t| \sqrt{2 \Lambda_\mathbf{k} + 3},
\end{align}
which are independent of the spin of the particle.
Here the first band is flat and non-dispersive in $\mathbf{k}$ space, and the dispersive 
bands are characterized by 
$
\Lambda_\mathbf{k} = \sum_{i = 1}^3 \cos(\mathbf{k} \cdot \mathbf{a}_i).
$
We choose a negative $t = -|t|$ in this paper leading to a flat band at the bottom. 
The resultant band structure is shown in Fig.~\ref{fig:bs}(c). We note that while the flat 
band is in quadratic touch with a dispersive band at $\mathbf{k = 0}$, the dispersive 
bands touch each other and form Dirac cones at the six corners of the Brillouin zone.

Given our interest in the dispersion of the two-body bound states, we need not only 
the band structure but also the associated Bloch states $|n \mathbf{k} \sigma \rangle$. 
Thus a compact way to express the eigenvectors of $\mathbf{H}_{\mathbf{k} \sigma}$ 
is~\cite{barreteau17, mizoguchi19}
\begin{align}
\label{eqn:1k}
|1 \mathbf{k} \sigma \rangle \equiv
\begin{pmatrix}
1_{A \mathbf{k} \sigma} \\ 1_{B \mathbf{k} \sigma} \\ 1_{C \mathbf{k} \sigma}
\end{pmatrix} 
&= \mathcal{A}_{1\mathbf{k}}
\begin{bmatrix}
\sin(\theta_{2\mathbf{k}} - \theta_{3\mathbf{k}}) \\
\sin(\theta_{3\mathbf{k}} - \theta_{1\mathbf{k}}) \\
\sin(\theta_{1\mathbf{k}} - \theta_{2\mathbf{k}})
\end{bmatrix}, \\
\label{eqn:2k}
|2 \mathbf{k} \sigma \rangle \equiv
\begin{pmatrix}
2_{A \mathbf{k} \sigma} \\ 2_{B \mathbf{k} \sigma} \\ 2_{C \mathbf{k} \sigma}
\end{pmatrix} 
&= \mathcal{A}_{2\mathbf{k}}
\begin{bmatrix}
\sin(\theta_{1\mathbf{k}} + \phi_\mathbf{k}) \\
\sin(\theta_{2\mathbf{k}} + \phi_\mathbf{k}) \\
\sin(\theta_{3\mathbf{k}} + \phi_\mathbf{k})
\end{bmatrix}, \\
\label{eqn:3k}
|3 \mathbf{k} \sigma \rangle \equiv
\begin{pmatrix}
3_{A \mathbf{k} \sigma} \\ 3_{B \mathbf{k} \sigma} \\ 3_{C \mathbf{k} \sigma}
\end{pmatrix} 
&= \mathcal{A}_{3\mathbf{k}}
\begin{bmatrix}
\cos(\theta_{1\mathbf{k}} + \phi_\mathbf{k}) \\
\cos(\theta_{2\mathbf{k}} + \phi_\mathbf{k}) \\
\cos(\theta_{3\mathbf{k}} + \phi_\mathbf{k})
\end{bmatrix},
\end{align}
where $n_{S \mathbf{k} \sigma} = \langle S |n \mathbf{k} \sigma \rangle$ is the projection 
of the Bloch state onto the $S$th sublattice, $\mathcal{A}_{n\mathbf{k}}$ is the 
normalization factor, and
$
\theta_{1\mathbf{k}} = k_xa/4 + k_ya/(4\sqrt{3}),
$
$
\theta_{2\mathbf{k}} = - k_ya/(2\sqrt{3}),
$
$
\theta_{3\mathbf{k}} = - k_xa/4 + k_ya/(4\sqrt{3})
$
and
$
\phi_\mathbf{k} = \frac{1}{2} \mathrm{arg}[e^{ik_y a/\sqrt{3}} + 2\cos(k_x a/2)e^{-ik_y/(2\sqrt{3})}]
$
are some phase factors associated with the geometry of the Kagome lattice.
Having completed the analysis of the one-body problem, next we proceed with 
the two-body problem.

\subsection{Two-body bound states}
\label{sec:tbs}

In the presence of a multiband tight-binding lattice, it is possible to solve the two-body 
problem exactly through a variational approach that is based on the following 
ansatz~\cite{iskin21}
\begin{align}
| \Psi_\mathbf{q} \rangle = \sum_{nm \mathbf{k}}
\alpha_{nm \mathbf{k}}^\mathbf{q} c_{n, \mathbf{k}+\mathbf{q}/2, \uparrow}^\dagger
c_{m, -\mathbf{k}+\mathbf{q}/2, \downarrow}^\dagger | 0 \rangle.
\end{align}
Here $\mathbf{q}$ is the center-of-mass momentum of the spin-singlet bound pair that 
is formed between a spin-$\uparrow$ and a spin-$\downarrow$ particle, the variational 
parameter $\alpha_{nm \mathbf{k}}^\mathbf{q}$ is a complex number in general, 
and the operator $c_{n \mathbf{k} \sigma}^\dagger$ creates a particle in the Bloch 
state 
$
|n\mathbf{k} \sigma \rangle = c_{n \mathbf{k} \sigma}^\dagger | 0 \rangle.
$
The creation operators in the orbital and Bloch basis are related through
$
c_{n \mathbf{k} \sigma}^\dagger = \sum_S n_{S \mathbf{k} \sigma} 
c_{S \mathbf{k} \sigma}^\dagger
$
since
$
\sum_S |S \mathbf{k} \sigma \rangle \langle S \mathbf{k} \sigma| = \mathcal{I}_{N_b}
$ 
is an identity operator in $N_b$ dimensions for a given $\mathbf{k}$ and $\sigma$.
The dispersion $E_{\ell \mathbf{q}}$ of the bound state is determined through the 
minimization of 
$
\langle \Psi_\mathbf{q} | \mathcal{H} - E_{\ell \mathbf{q}} | \Psi_\mathbf{q} \rangle = 0
$
with respect to $\alpha_{nm \mathbf{k}}^\mathbf{q}$, where 
$
\mathcal{H}  = \mathcal{H}_\uparrow + \mathcal{H}_\downarrow 
+ \mathcal{H}_{\uparrow\downarrow}
$
is the total Hamiltonian of the system. 
Here we limit our analysis to an onsite (i.e., contact) attraction between the 
particles thanks mainly to the clarity of the central theme of this paper. 
For this purpose let us consider
$
\mathcal{H}_{\uparrow\downarrow} = - U \sum_{S i} \rho_{S i \uparrow} \rho_{S i \downarrow},
$
where $U \ge 0$ is the strength of the interaction and
$
\rho_{S i \sigma} = c_{S i \sigma}^\dagger c_{S i \sigma}
$
is the number operator at the $S$th sublattice in the $i$th unit cell. 
Here the operator $c_{S i \sigma}$ 
corresponds to the Fourier transform of $c_{S \mathbf{k} \sigma}$.
In addition we take advantage of the time-reversal symmetry, and set
$
n_{S \mathbf{k} \uparrow} = n_{S, -\mathbf{k}, \downarrow}^* \equiv n_{S \mathbf{k}}
$
and
$
\varepsilon_{n \mathbf{k} \uparrow} = \varepsilon_{n, -\mathbf{k}, \downarrow} 
\equiv \varepsilon_{n \mathbf{k}}.
$
After some straightforward algebra~\cite{iskin21}, $E_{\ell \mathbf{q}}$ is characterized 
by a set of linear equations
\begin{equation}
\label{eqn:GB}
\mathbf{G^q} \boldsymbol{\beta}_{\ell \mathbf{q}} = 0,
\end{equation}
where $\mathbf{G^q}$ is an $N_b$-dimensional Hermitian matrix with the following 
elements
\begin{align}
G_{SS'}^\mathbf{q} = \delta_{SS'} - \frac{U}{N_c} \sum_{n m S' \mathbf{k}}
\frac{m_{S \mathbf{K'}}^* n_{S \mathbf{K}} n_{S' \mathbf{K}}^* m_{S' \mathbf{K'}}}
{\varepsilon_{n \mathbf{K}} + \varepsilon_{m \mathbf{K'}} - E_{\ell \mathbf{q}}}.
\label{eqn:FSS}
\end{align}
Here 
$
\mathbf{K} = \mathbf{k}+\mathbf{q}/2
$
and
$
\mathbf{K'} = \mathbf{k}-\mathbf{q}/2.
$
Thus $E_{\ell \mathbf{q}}$ is determined by setting $\det \mathbf{G^q} = 0$, and there 
are $N_b$ solutions for a given $\mathbf{q}$. We label these solutions with $\ell = \{1,2,3\}$
starting from the lower branch. Furthermore the state vector
$
\boldsymbol{\beta}_{\ell \mathbf{q}} = 
(\beta_{A \ell \mathbf{q}}\; \beta_{B \ell \mathbf{q}}\; \beta_{C \ell \mathbf{q}})^\mathrm{T}
$
with
$
\beta_{S \ell \mathbf{q}} = \sum_{nm\mathbf{k}} \alpha_{nm \mathbf{k}}^\mathbf{q} 
n_{S \mathbf{K}} m_{S \mathbf{K'}}^*
$
is the corresponding eigenvector of $\mathbf{G^q}$, and it carries further insight into 
the physical mechanism and nature of the bound state.

\begin{widetext}
\begin{center}
\begin{figure} [htb]
\centerline{\scalebox{0.45}{\includegraphics{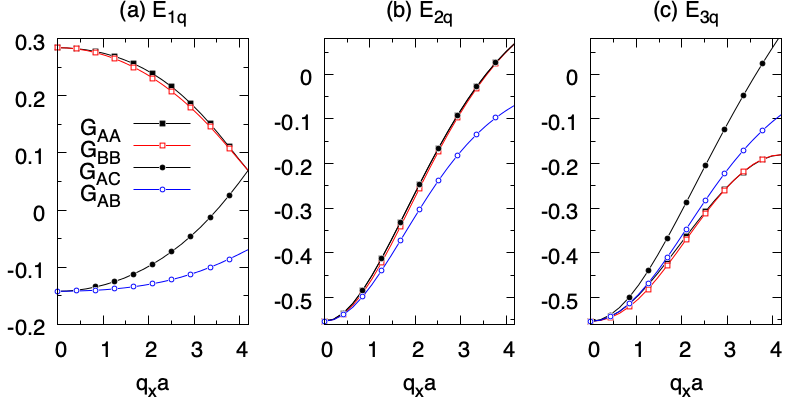}}}
\caption{\label{fig:G}
Matrix elements $G_{SS'}^\mathbf{q}$ for $U = 2|t|$ as a function of $q_x$ 
when $q_y = 0$. Here (a), (b) and (c) corresponds, respectively, to the self-consistent 
solutions for $E_{1 \mathbf{q}}$, $E_{2 \mathbf{q}}$ and $E_{3 \mathbf{q}}$.
Since 
$
G_{CC}^\mathbf{q} = G_{AA}^\mathbf{q}
$
and
$
G_{BC}^\mathbf{q} = G_{AB}^\mathbf{q}
$
for all $q_x$, these coefficients are not shown. Note that
$
G_{AA}^\mathbf{q} = G_{BB}^\mathbf{q}
$
and
$
G_{AB}^\mathbf{q} = G_{AC}^\mathbf{q}
$
in the small-$q_x a$ limit. In addition, together with the characteristic (bound-state) 
Eq.~(\ref{eqn:GB}), these coefficients imply $\beta_{1 \mathbf{q}} \propto (1,1,1)$, 
$\beta_{2 \mathbf{q}} \propto (1,0,-1)$ and $\beta_{3 \mathbf{q}} \propto (1,-2,1)$ 
in the small-$q_x a$ limit.
}
\end{figure}
\end{center}
\end{widetext}

Our numerical calculations for the Kagome lattice show that the matrix elements of 
$\mathbf{G^q}$ have the following properties. When $\mathbf{q}$ is along one of the 
principal axis, i.e., either when $q_x = 0$ or $q_y = 0$, we observed that
$
G_{AA}^\mathbf{q} = G_{CC}^\mathbf{q}
$
and
$
G_{AB}^\mathbf{q} = G_{BC}^\mathbf{q}
$
for all parameters. Thus one of the eigenvalues of $\mathbf{G^q} = 0$ is
$
G_{AA}^\mathbf{q} - G_{AC}^\mathbf{q}
$ 
with the eigenvector
$
\beta_{2 \mathbf{q}} \propto (1, 0, -1),
$
and it determins the middle branch $E_{2 \mathbf{q}}$.
The upper branch $E_{3 \mathbf{q}}$ and the lower one $E_{1 \mathbf{q}}$ are determined,
respectively, by the eigenvalues
$
(G_{AA}^\mathbf{q}+ G_{BB}^\mathbf{q} + G_{AC}^\mathbf{q})/2 \pm 
[8(G_{AA}^\mathbf{q})^2 + (G_{AC}^\mathbf{q})^2+(G_{AA}^\mathbf{q}-G_{BB}^\mathbf{q})
(2G_{AC}^\mathbf{q}+G_{AA}^\mathbf{q}-G_{BB}^\mathbf{q})]^{1/2}/2,
$
where
$
\beta_{3 \mathbf{q}} \propto (1, Q_\mathbf{q}, 1)
$
with $Q_\mathbf{q} < 0$ and
$
\beta_{1 \mathbf{q}} \propto (1, R_\mathbf{q}, 1)
$
with $R_\mathbf{q} > 0$. While the $\mathbf{q}$ dependences of $Q_\mathbf{q}$ and 
$R_\mathbf{q}$ are not very illuminating and skipped, they are in such a way that 
$Q_\mathbf{q} R_\mathbf{q} = -2$ for every $\mathbf{q}$ which is required by the 
orthonormalization of $\boldsymbol{\beta}_{\ell \mathbf{q}}$. Thus, by setting the eigenvalues 
of $\mathbf{G^q}$ to 0, we find that $E_{2 \mathbf{q}}$ is determined by the condition
$
G_{AA}^\mathbf{q} = G_{AC}^\mathbf{q},
$
and that $E_{1 \mathbf{q}}$ and $E_{3 \mathbf{q}}$ are determined by the same condition
$
2(G_{AB}^\mathbf{q})^2 = G_{BB}^\mathbf{q}(G_{AA}^\mathbf{q} + G_{AC}^\mathbf{q}).
$
On the other hand, when $q = \sqrt{q_x^2+q_y^2}$ is small, i.e., when $q a \ll 1$, we observed that
$
G_{AA}^\mathbf{q} = G_{BB}^\mathbf{q} = G_{CC}^\mathbf{q}
$
and
$
G_{AB}^\mathbf{q} = G_{BC}^\mathbf{q} = G_{AC}^\mathbf{q}
$
for all parameters. These are shown in Fig.~\ref{fig:G}.
Thus $E_{1 \mathbf{q}}$ is determined by the condition
$
G_{AA}^\mathbf{q} = - 2G_{AB}^\mathbf{q},
$
and it is characterized by $\beta_{1 \mathbf{q}} \propto (1,1,1)$. This suggests 
that the low-energy bound states can be distinguished by their perfectly in-phase 
(i.e., spatially-uniform) contribution from all three sublattices~\cite{twobandnote}. 
On the other hand $E_{2 \mathbf{q}}$ and $E_{3 \mathbf{q}}$ are both determined by 
the very same condition
$
G_{AA}^\mathbf{q} = G_{AB}^\mathbf{q}
$
(i.e., they are degenerate in the small-$q$ limit), and are characterized, respectively, by 
$\beta_{2 \mathbf{q}} \propto (1,0,-1)$ and $\beta_{3 \mathbf{q}} \propto (1,-2,1)$.

\begin{figure} [htb]
\centerline{\scalebox{0.5}{\includegraphics{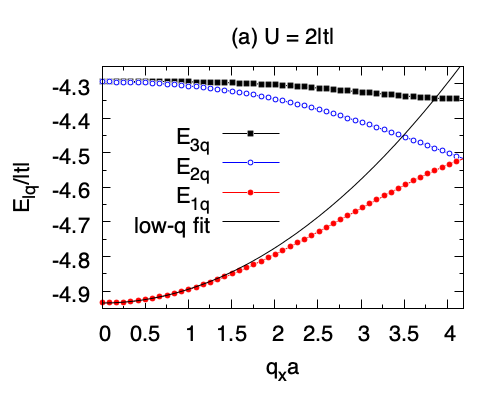}}}
\centerline{\scalebox{0.45}{\includegraphics{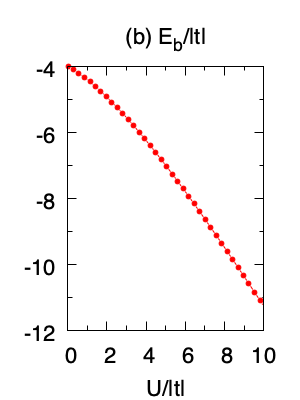} \includegraphics{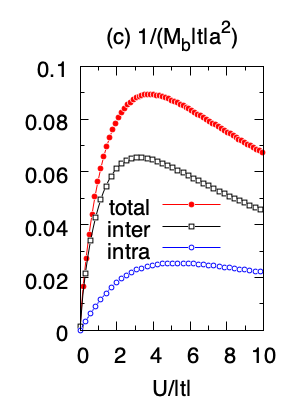}}}
\caption{\label{fig:tb} 
(a) Two-body spectrum $E_{\ell \mathbf{q}}$ for $U = 2|t|$ as a function of $q_x$ when 
$q_y = 0$. The quadratic expansion of $E_{1 \mathbf{q}} = E_b + q^2/(2M_b)$ is an 
excellent fit for the lower branch in the small-$q$ limit.
(b) $E_b$ as a function of $U$. 
(c) $1/M_b = 1/M_b^\mathrm{intra} + 1/M_b^\mathrm{inter}$ as a function of $U$ along 
with its intraband and interband contributions,
where
$
1/M_b^\mathrm{inter} = 1/M_b^\mathrm{inter,1} + 1/M_b^\mathrm{inter,2}.
$
}
\end{figure}

As an illustration we set $U = 2|t|$ and $q_y = 0$ in Fig.~\ref{fig:tb}(a), and present the 
two-body spectrum $E_{\ell \mathbf{q}}$ as a function of $q_x$. Even though $E_{3 \mathbf{q}}$
disperses in $\mathbf{q}$, it appears quite flat and featureless in the presented scale.
We observe that while $E_{3 \mathbf{q}}$ and $E_{2 \mathbf{q}}$ are degenerate at low $q$, 
$E_{2 \mathbf{q}}$ and $E_{1 \mathbf{q}}$ are degenerate at the corner of the Brillouin zone. 
More importantly the quadratic expansion of $E_{1 \mathbf{q}} = E_b + q^2/(2M_b)$ provides 
an excellent fit for the lower branch in the small-$q$ limit, where $E_b \approx  -4.934|t|$ is the 
energy of the lowest bound state and $M_b \approx 12.574/(|t|a^2)$ is its effective mass. 
We remark here that the effective masses of $E_{1 \mathbf{q}}$ and $E_{2 \mathbf{q}}$ 
branches become very close to each other (in magnitude) as $U$ gets larger and larger. 
In order to gain deeper insight into the former result, next we use our observation that 
$
\beta_{1 \mathbf{q}} \propto (1,1,1) 
$
for the lower band in the small-$q$ limit. Furthermore this observation allows us to derive the 
generalized relation between the inverse of the effective-mass tensor of the lowest bound 
states and the quantum-metric tensor of the underlying Bloch states for those multiband 
lattices that simultaneously exhibit time-reversal symmetry and fulfill the condition on 
spatially-uniform pairing.

\section{Relation to quantum metric}
\label{sec:qm}

Given our numerical observation that the small-$q$ limit of the lower branch 
$E_{1 \mathbf{q}}$ is characterized by 
$
\beta_{1 \mathbf{q}} \propto (1,1,1),
$ 
it is possible to isolate the condition that determines $E_{1 \mathbf{q}}$ as
$
\sum_{SS'} G_{SS'}^\mathbf{q} = 0.
$
This is a very convenient form, and it may find practical applications in other 
multiband lattices as long as the lowest bound states are distinguished by a 
perfectly in-phase contribution from all of the sublattices, i.e., 
$
\beta_{1 \mathbf{q}} \propto (1,1, \dots, 1).
$ 
Hoping that this is generically the case in time-reversal-symmetric systems with a
spatially-uniform pairing, below we keep the formalism and the discussion general.
By plugging Eq.~(\ref{eqn:FSS}) into this condition, and observing that
$
\sum_S m_{S \mathbf{K'}}^* n_{S \mathbf{K}}  = \langle m_\mathbf{K'}| n_\mathbf{K} \rangle
$
with $|n_\mathbf{K} \rangle$ representing the Bloch states that are given in 
Eqs.~(\ref{eqn:1k}-\ref{eqn:3k}), we obtain a much simpler condition
\begin{align}
\label{eqn:sce}
1 = \frac{U}{N} \sum_{n m \mathbf{k}}
\frac{|\langle m_{\mathbf{k}-\mathbf{q}/2} | n_{\mathbf{k}+\mathbf{q}/2} \rangle|^2}
{\varepsilon_{n, \mathbf{k}+\mathbf{q}/2} + \varepsilon_{m, \mathbf{k}-\mathbf{q}/2} - E_{1 \mathbf{q}}}.
\end{align}
Here $N = N_b N_c$ is the total number of lattice sites in the system. This expression 
can be used to make further analytical progress by taking its small-$q$ limit. 

For instance, in the presence of an energetically-isolated flat band
$
\varepsilon_{f\mathbf{k}} = \varepsilon_f
$ 
that is separated from the remaining bands with a finite band gap, Eq.~(\ref{eqn:sce}) 
can be approximated by
$
1 = (U/N) \sum_\mathbf{k}
|\langle f_{\mathbf{k}-\mathbf{q}/2} | f_{\mathbf{k}+\mathbf{q}/2} \rangle|^2
/ (2\varepsilon_f - E_{1 \mathbf{q}})
$
in the small-$U$ limit, leading to~\cite{torma18}
$
E_{1 \mathbf{q}} = 2\varepsilon_f  
- (U/N) \sum_\mathbf{k} |\langle f_{\mathbf{k}-\mathbf{q}/2} | f_{\mathbf{k}+\mathbf{q}/2} \rangle|^2.
$
Furthermore we use the Taylor expansions
$
|f_{\mathbf{k} \pm \mathbf{q}/2} \rangle = |f_\mathbf{k} \pm (1/2) \sum_i q_i \partial_i f_\mathbf{k}
+ (1/8) \sum_{ij} q_i q_j \partial_i \partial_j f_\mathbf{k} \rangle
$
in small $\mathbf{q}$ and obtain
$
|\langle f_{\mathbf{k}-\mathbf{q}/2} | f_{\mathbf{k}+\mathbf{q}/2} \rangle|^2 = 
1 - (1/2) \sum_{ij} q_i q_j g_{ij}^{f \mathbf{k}}.
$
This is exact up to second order in $\mathbf{q}$ where
$
g_{ij}^{n\mathbf{k}} = \mathrm{Tr} \big( \partial_i P_{n_\mathbf{k}} \partial_j P_{n_\mathbf{k}} \big)
$
is the matrix element of the so-called quantum-metric tensor of the Bloch state 
$|n_\mathbf{k} \rangle$~\cite{resta11}. Here $\mathrm{Tr}$ is the trace, 
$\partial_i = \partial/\partial k_i$ is the partial derivative, and
$
P_{n_\mathbf{k}} = | n_\mathbf{k} \rangle \langle n_\mathbf{k} |
$
is the projection operator. Since $P_{n_\mathbf{k}}$ is a gauge-independent operator by 
definition, $g_{ij}^{n\mathbf{k}}$ is also a band invariant. By noting that
$
\langle n_\mathbf{k} | \partial_i n_\mathbf{k} \rangle = - \langle \partial_i n_\mathbf{k} | n_\mathbf{k} \rangle 
$
is an imaginary number due to the normalization
$
\langle n_\mathbf{k} | n_\mathbf{k} \rangle = 1
$ 
of the Bloch states, the quantum metric can be reexpressed in a more familiar 
form~\cite{provost80, berry89, resta11}
\begin{align}
\label{eqn:gijn}
g_{ij}^{n\mathbf{k}} = 2\mathrm{Re} \langle \partial_i n_\mathbf{k} |
(\mathcal{I}_{N_b} - | n_\mathbf{k} \rangle \langle n_\mathbf{k} |)
| \partial_j n_\mathbf{k} \rangle,
\end{align}
where $\mathrm{Re}$ is the real part of the expression. As a result we eventually find
(i.e., to the lowest order in $\mathbf{q}$)
\begin{align}
\label{eqn:E1q}
E_{1 \mathbf{q}} = E_b + \frac{1}{2} \sum_{ij} q_i q_j (M_b^{-1})_{ij},
\end{align}
where 
$
E_b = 2\varepsilon_f - U/N_b
$
is the threshold energy and
$
(M_b^{-1})_{ij} = (U/N) \sum_\mathbf{k} g_{ij}^{f \mathbf{k}}
$
is the matrix element of the inverse of the effective-mass tensor $\mathbf{M_b}$. 
We remark here that these expressions are strictly valid in the small-$U$ limit 
assuming an energetically-isolated flat band~\cite{torma18}.

Similarly one can perform a small-$\mathbf{q}$ expansion of Eq.~(\ref{eqn:sce}) and 
generalize $E_b$ and $\mathbf{M}_\mathbf{b}^{-1}$ not only to arbitrary $U$ values but 
also to arbitrary band structures. For this purpose we first use the Taylor expansions of 
$|n_{\mathbf{k} \pm \mathbf{q}/2} \rangle$ given above and find
$
\langle m_{\mathbf{k}-\mathbf{q}/2} | n_{\mathbf{k}+\mathbf{q}/2} \rangle = 
\delta_{mn} + \sum_i q_i \langle m_\mathbf{k} | \partial_i n_\mathbf{k} \rangle
 - (1/8) \sum_{ij} q_i q_j (3 \langle \partial_i m_\mathbf{k} | \partial_j n_\mathbf{k} \rangle 
 +  \langle \partial_j m_\mathbf{k} | \partial_i n_\mathbf{k} \rangle ),
$
where 
$
\langle m_\mathbf{k} | n_\mathbf{k} \rangle = \delta_{mn}
$
is due to the orthonormalization of the Bloch states and
$
\langle m_\mathbf{k} | \partial_i n_\mathbf{k} \rangle = - \langle \partial_i m_\mathbf{k} | n_\mathbf{k} \rangle.
$
This leads to
$
|\langle m_{\mathbf{k}-\mathbf{q}/2} | n_{\mathbf{k}+\mathbf{q}/2} \rangle|^2 = 
\delta_{mn} - \mathrm{Re} \sum_{ij} q_iq_j 
\langle \partial_i n_\mathbf{k}|
(\delta_{mn} - |m_\mathbf{k} \rangle \langle m_\mathbf{k}|)
|\partial_j n_\mathbf{k} \rangle
$
in the small-$\mathbf{q}$ limit. Then we expand the single-particle spectrum
$
\varepsilon_{n, \mathbf{k} \pm \mathbf{q}/2} = \varepsilon_{n \mathbf{k}} 
\pm (1/2) \sum_i q_i \partial_i\varepsilon_{n \mathbf{k}}
+ (1/8) \sum_{ij} q_i q_j \partial_i \partial_j \varepsilon_{n \mathbf{k}}
$
up to second order in $\mathbf{q}$ and plug Eq.~(\ref{eqn:E1q}) for the dispersion of 
the lowest bound states. By matching the coefficient of the zeroth order terms in
Eq.~(\ref{eqn:sce}), we find
\begin{align}
\label{eqn:Eb}
1 = \frac{U}{N} \sum_{n \mathbf{k}} \frac{1}{2\varepsilon_{n \mathbf{k}} - E_b},
\end{align}
which is the self-consistency relation for the energy $E_b$ of the lowest bound state. 
The first-order terms already vanish. By requiring that the second-order terms vanish, 
we find a closed-form expression for 
$
\mathbf{M}_\mathbf{b}^{-1} = \mathbf{M}_\mathrm{intra}^{-1} + \mathbf{M}_\mathrm{inter}^{-1},
$
where $\mathbf{M}_\mathrm{intra}^{-1}$ is the so-called conventional or the intraband and
$
\mathbf{M}_\mathrm{inter}^{-1} = \mathbf{M}_\mathrm{inter, 1}^{-1} + \mathbf{M}_\mathrm{inter, 2}^{-1}
$ 
is the so-called geometric or the interband contribution to the inverse of the effective-mass tensor.
They can be written as
\begin{align}
\label{eqn:intra}
(M_\mathrm{intra}^{-1})_{ij} &= \frac{1}{2D}
\sum_{n \mathbf{k}} \frac{\partial_i \partial_j \varepsilon_{n \mathbf{k}}}{(2\varepsilon_{n \mathbf{k}} - E_b)^2}, \\
\label{eqn:inter1}
(M_\mathrm{inter, 1}^{-1})_{ij} &= \frac{1}{D}
\sum_{n \mathbf{k}} \frac{g_{ij}^{n \mathbf{k}}}{2\varepsilon_{n \mathbf{k}} - E_b}, \\
\label{eqn:inter2}
(M_\mathrm{inter, 2}^{-1})_{ij} &= - \frac{1}{D} 
\sum_{n, m \ne n, \mathbf{k}} \frac{ g_{ij}^{nm\mathbf{k}}  }
{\varepsilon_{n \mathbf{k}} + \varepsilon_{m \mathbf{k}} - E_b},
\end{align}
where 
$
D = \sum_{n \mathbf{k}} \frac{1}{(2\varepsilon_{n \mathbf{k}} - E_b)^2}
$
and $E_b$ is determined by Eq.~(\ref{eqn:Eb}) for a given $U$. Here 
Eq.~(\ref{eqn:inter2}) depends on the so-called band-resolved quantum metric
\begin{align}
g_{ij}^{nm\mathbf{k}} = 2\mathrm{Re} \langle \partial_i n_\mathbf{k} | m_\mathbf{k} \rangle
\langle m_\mathbf{k} | \partial_j n_\mathbf{k} \rangle,
\end{align}
since it produces the quantum metric of the $n$th band when summed over the 
rest of the bands, i.e.,
$
g_{ij}^{n\mathbf{k}} = \sum_{m \ne n} g_{ij}^{nm\mathbf{k}}. 
$
Equations~(\ref{eqn:intra} -~\ref{eqn:inter2}) constitute the generalized relation 
between the inverse of the effective-mass tensor of the lowest bound states and the 
quantum-metric tensor of the underlying Bloch states, and they are exact. 

Let us now reproduce the known results using Eqs.~(\ref{eqn:Eb} -~\ref{eqn:inter2}).
In the case of an energetically-isolated flat band
$
\varepsilon_{f \mathbf{k}} = \varepsilon_f
$ 
that is separated from the remaining bands with a finite band gap, Eqs.~(\ref{eqn:intra}) 
and~(\ref{eqn:inter2}) are negligible in the small-$U$ limit. Furthermore Eq.~(\ref{eqn:inter1}) 
is approximated by
$
(M_\mathrm{inter, 1}^{-1})_{ij} = (U/N) \sum_\mathbf{k} g_{ij}^{f \mathbf{k}}
$
in the small-$U$ limit where 
$
2\varepsilon_f  - E_b = U/N_b.
$
These are in full agreement with the literature~\cite{torma18} and the discussion 
given below Eq.~(\ref{eqn:sce}).
On the other hand, in the case of a two-band ($N_b = 2$) lattice that is described 
by the Hamiltonian density
$
H_{\mathbf{k} \sigma} = d_0^\mathbf{k} \mathcal{I}_2 + \mathbf{d}_\mathbf{k} \cdot \boldsymbol{\tau},
$
the single-particle spectrum is given by
$
\varepsilon_{s \mathbf{k}} = d_0^\mathbf{k} + s d_\mathbf{k}
$
where $s = \{+, -\}$ labels, respectively, the upper and lower bands. Here 
$\boldsymbol{\tau} = (\tau_x, \tau_y, \tau_z)$ is a vector of Pauli matrices in the 
two-dimensional orbital basis. In this case the quantum metrics of the two bands 
are equal to each other, i.e., 
$
g_{ij}^{+, \mathbf{k}}  = g_{ij}^{-, \mathbf{k}},
$ 
given that
$
g_{ij}^{s \mathbf{k}} = 2\mathrm{Re} 
\langle \partial_i s_\mathbf{k} | 
(-s)_\mathbf{k} \rangle \langle (-s)_\mathbf{k} 
| \partial_j s_\mathbf{k} \rangle
$
and
$
\sum_{s \mathbf{k}} |s_\mathbf{k} \rangle \langle s_\mathbf{k} | = \mathcal{I}_2.
$
For this reason Eq.~(\ref{eqn:inter2}) can be written as 
$
\sum_{s \mathbf{k}} g_{ij}^{s\mathbf{k}} / (2d_0^\mathbf{k} - E_b),
$
leading to
$
(M_\mathrm{inter}^{-1})_{ij} =  -\frac{2}{D} \sum_{s \mathbf{k}}\frac{s d_\mathbf{k} g_{ij}^{s \mathbf{k}}}
{(2\varepsilon_{s \mathbf{k}} - E_b)(2d_0^\mathbf{k} - E_b)}.
$
These are again in full agreement with the literature~\cite{iskin21}.

Lastly we apply Eqs.~(\ref{eqn:Eb} -~\ref{eqn:inter2}) to the Kagome lattice and 
solve them self-consistently for $E_b$ and $M_b$. Their numerical values are 
presented, respectively, in Figs.~\ref{fig:tb}(b) and~\ref{fig:tb}(c) as a function of $U$. 
In this particular case $\mathbf{M_b}$ is an isotropic matrix with 
$
(M_b^{-1})_{ij} = \delta_{ij}/M_b.
$
We find that while $E_b = -4|t| - U/3$ increases linearly in the small-$U$ limit due to
the presence of a flat lower band in a three-band lattice, $E_b = - U$ in the large-$U$ 
limit which is similar to what happens in a one-band lattice. 
Similarly while $M_b = r_1/[a^2 U \ln(r_2|t|/U)]$ diverges logarithmically in the small-$U$ 
limit due to the presence of a band touching with a non-isolated flat band, 
$M_b = U/(a^2 t^2)$ increases linearly in the large-$U$ limit which is again 
similar to what happens in a one-band lattice. Here $r_1$ and $r_2$ are real positive 
constants. In comparison, in the absence of a band touching (i.e., for an isolated flat band), 
we note that $M_b = r_3/(a^2 U)$ diverges with a power law in the small-$U$ limit~\cite{torma18}.
Furthermore we find that $E_b \approx  -4.934|t|$ and $M_b \approx 12.574/(|t|a^2)$ 
when $U = 2|t|$, and they provide a perfect fit for the exact results in the small-$q$ limit. 
This is shown in Fig.~\ref{fig:tb}(a). 

We note in passing that $1/M_b^\mathrm{inter,1} > 0$ is in direct competition with 
$1/M_b^\mathrm{inter,2} < 0$, and their magnitudes are about three orders of 
magnitude larger than $1/M_b^\mathrm{intra}$. However their sum 
$
1/M_b^\mathrm{inter} = 1/M_b^\mathrm{inter,1} + 1/M_b^\mathrm{inter,2}
$ 
is quite comparable to $1/M_b^\mathrm{intra}$ as can be seen in Fig.~\ref{fig:tb}(c). 
Having shown that Eqs.~(\ref{eqn:intra}-\ref{eqn:inter2}) are exact for the Kagome lattice
for all $U$ values, next we discuss their versatility for other lattices.

\section{Spatially-Uniform Pairing}
\label{sec:up}

In accordance with the analysis presented above, Eqs.~(\ref{eqn:intra}-\ref{eqn:inter2}) 
are clearly exact for those multiband lattices that simultaneously fulfill the following 
conditions: 
(i) the Bloch Hamiltonian must exhibit time-reversal symmetry and 
(ii) the resultant two-body wave function must have a uniform contribution from 
all of the underlying sublattices.
The latter is the so-called spatially-uniform-pairing condition, and it is expected to be 
satisfied by those Bloch Hamiltonians that are invariant under the interchange of their 
sublattices. Note that if the condition (ii) is satisfied for the two-body problem in 
a lattice then we expect the mean-field pairing order parameter ($\Delta_{Si} = \Delta_S$) 
for the many-body problem to be spatially-uniform, i.e., $\Delta_S = \Delta$ is equal 
for all of the sublattices. 

In the case of two-band lattices while the honeycomb, Mielke checkerboard, Kane-Mele, 
Creutz and Haldane type Hubbard models with onsite interactions are among those
popular lattices that satisfy condition (ii), i.e., because of their inversion symmetry, 
the sawtooth and zigzag type models are not. Here we note that the time-reversal 
symmetry is broken for the Creutz and Haldane models. 
In the case of three-band lattices, while the Kagome lattice with onsite interaction 
satisfy condition (ii), the Lieb and dice lattices do not since only two of their 
sublattices are interchangeable with each other but not the third one. 
According to the recent findings~\cite{torma18, wu21}, the contribution to the 
two-body wave function from the non-interchangeable sublattice vanishes for 
both of these models in the small-$U$ limit.
Because of this Eqs.~(\ref{eqn:intra}-\ref{eqn:inter2}) still work for these models
but only in the $U/t \to 0$ limit. 
In particular, since the flat band of the Lieb lattice is isolated from the other bands 
with a gap, it is sufficient to keep only the flat-band contribution coming from 
Eq.~(\ref{eqn:inter1}) in the small-$U$ limit.
However, since the flat band of the dice lattice is in touch with one of the dispersive 
bands, one needs to keep both the flat-band contribution and that of the touching band 
coming from Eqs.~(\ref{eqn:intra}-\ref{eqn:inter2}) in the small-$U$ limit.

\section{Conclusion}
\label{sec:conc}

To summarize here we considered an onsite attraction $U$ between a spin-$\uparrow$ 
and a spin-$\downarrow$ fermion in a multiband lattice, and derived an exact relation 
between the inverse of the effective-mass tensor $(M_b^{-1})_{ij}$ of the lowest bound 
states $E_{1 \mathbf{q}}$ and the quantum-metric tensor $g_{ij}^{n\mathbf{k}}$ of the 
underlying Bloch states $|n_\mathbf{k} \rangle$. In addition to the intraband contribution 
$(M_\mathrm{intra}^{-1})_{ij}$ that depends only on the single-particle spectrum 
$\varepsilon_{n\mathbf{k}}$ and the interband contribution $(M_\mathrm{inter,1}^{-1})_{ij}$ 
that is controlled by $g_{ij}^{n\mathbf{k}}$, our generalized relation has an additional 
interband contribution $(M_\mathrm{inter,2}^{-1})_{ij}$ that depends on the 
band-resolved quantum metric $g_{ij}^{nm\mathbf{k}}$. 
Our analytical expression is applicable to those multiband lattices that simultaneously 
exhibit time-reversal symmetry and fulfill the condition on spatially-uniform pairing.
It reproduces the previously known results including that of isolated flat bands in the 
small-$U$ limit~\cite{torma18} and that of two-band 
lattices for arbitrary $U$~\cite{iskin21}. Furthermore we also solved the two-body 
problem in a Kagome lattice with nearest-neighbor hoppings, and showed 
that the exact relation provides a perfect benchmark for this three-band lattice.
In general it is probably not possible to isolate the geometric contributions to the 
effective mass, and study their effects alone in the experiments. However our results 
for a flat band show that the geometric contributions play a dominant role in the 
small-$U$ limit, and can be studied there.
As an outlook our exact relation may find direct applications in many other lattices 
including those of the Moire materials~\cite{topp21}, motivated by the hope that the 
formation of a two-body bound state can be used as a precursor to superconductivity 
in these systems.

\begin{acknowledgments}
The author acknowledges funding from T{\"U}B{\.I}TAK Grant No. 1001-118F359.
\end{acknowledgments}


\begin{thebibliography}{99}

\bibitem{provost80}
J. P. Provost and G. Vallee,
Riemannian structure on manifolds of quantum states,
Commun. Math. Phys. \textbf{76}, 289 (1980).

\bibitem{berry89}
M. V. Berry,
The quantum phase, five years after in Geometric Phases in Physics,
edited by A. Shapere and F. Wilczek (World Scientific, Singapore, 1989).

\bibitem{resta11}
R. Resta, 
The insulating state of matter: A geometrical theory,
Eur. Phys. J. B \textbf{79}, 121 (2011).
%

\bibitem{peotta15}
S. Peotta and P. T\"{o}rm\"{a}, 
Superfluidity in topologically nontrivial flat bands,
Nat. Commun. \textbf{6}, 8944 (2015).

\bibitem{julku16}
A. Julku, S. Peotta, T. I. Vanhala, D.-H. Kim, and P. T\"orm\"a,
Geometric Origin of Superfluidity in the Lieb-Lattice Flat Band,
Phys. Rev. Lett. \textbf{117}, 045303 (2016).

\bibitem{torma17a}
L. Liang, T. I. Vanhala, S. Peotta, T. Siro, A. Harju, and P. T\"{o}rm\"{a},
Band geometry, Berry curvature, and superfluid weight,
Phys. Rev. B {\bf 95}, 024515 (2017).

\bibitem{iskin18}
M. Iskin, 
Quantum metric contribution to the pair mass in spin-orbit coupled Fermi superfluids,
Phys. Rev. A \textbf{97}, 033625 (2018).

\bibitem{iskin19}
M. Iskin,
Origin of flat-band superfluidity on the Mielke checkerboard lattice, 
Phys. Rev. A \textbf{99}, 053608 (2019).

\bibitem{wang20}
Z. Wang, G. Chaudhary, Q. Chen, and K. Levin,
Quantum geometric contributions to the BKT transition: Beyond mean field theory,
Phys. Rev. B \textbf{102}, 184504 (2020).

\bibitem{iskin20}
M. Iskin, 
Collective excitations of a BCS superfluid in the presence of two sublattices, 
Phys. Rev. A \textbf{101}, 053631 (2020).

\bibitem{wu21}
Y.-R. Wu, X.-F. Zhang, C.-F. Liu, W.-M. Liu, and Y.-C. Zhang,
Superfluid density and collective modes of fermion superfluid in dice lattice,
Sci. Rep. \textbf{11}, 13572 (2021).
%

 
\bibitem{asteria19}
L. Asteria, D. T. Tran, T. Ozawa, M. Tarnowski, B. S. Rem, N. FlŠschner, K. Sengstock, 
N. Goldman, and C. Weitenberg, 
Measuring quantized circular dichroism in ultracold topological matter, 
Nat. Phys. \textbf{15}, 449 (2019).

\bibitem{yu20} 
M. Yu, P. Yang, M. Gong, Q. Cao, Q. Lu, H. Liu, S. Zhang, M. B. Plenio, F. Jelezko, T. Ozawa, 
N. Goldman, S. Zhang, and J. Cai,
Experimental measurement of the quantum geometric tensor using coupled qubits in diamond, 
Natl. Sci. Rev. \textbf{7}, 254 (2020).

\bibitem{gianfrate20}
A. Gianfrate, O. Bleu, L. Dominici, V. Ardizzone, M. De Giorgi, D. Ballarini, G. Lerario, K. West, 
L. Pfeiffer, D. Solnyshkov, D. Sanvitto, and G. Malpuech, 
Measurement of the quantum geometric tensor and of the anomalous Hall drift, 
Nature \textbf{578}, 381 (2020).

\bibitem{tan21}
X. Tan, D.-W. Zhang, Z. Yang, J. Chu, Y.-Q. Zhu, D. Li, X. Yang, S. Song, Z. Han, Z. Li, 
Y. Dong, H.-F. Yu, H. Yan, S.-L. Zhu, and Y. Yu,
Experimental measurement of the quantum metric tensor and related topological phase 
transition with a superconducting qubit, 
Phys. Rev. Lett. \textbf{122}, 210401 (2019).

\bibitem{liao21}
Q. Liao, C. Leblanc, J. Ren, F. Li, Y. Li, D. Solnyshkov, G. Malpuech, J. Yao, and H. Fu,
Experimental Measurement of the Divergent Quantum Metric of an Exceptional Point,
Phys. Rev. Lett. \textbf{127}, 107402 (2021).
%

%
\bibitem{torma18} 
P. T\"{o}rm\"{a},  L. Liang, and S. Peotta,
Quantum metric and effective mass of a two-body bound state in a flat band,
Phys. Rev. B \textbf{98}, 220511(R) (2018).

\bibitem{iskin21}
M. Iskin,
Two-body problem in a multiband lattice and the role of quantum geometry,
Phys. Rev. A \textbf{103}, 053311 (2021).
%


\bibitem{jo12}
G.-B. Jo, J. Guzman, C. K. Thomas, P. Hosur, A. Vishwanath, and D. M. Stamper-Kurn,
Ultracold Atoms in a Tunable Optical Kagome Lattice,
Phys. Rev. Lett. \textbf{108}, 045305 (2012).

\bibitem{nakata12}
Y. Nakata, T. Okada, T. Nakanishi, and M. Kitano,
Observation of flat band for terahertz spoof plasmons in a metallic Kagom\'e lattice,
Phys. Rev. B \textbf{85}, 205128 (2012).

\bibitem{li18}
Z. Li, J. Zhuang, L. Wang, H. Feng, Q. Gao, X. Xu, W. Hao, X. Wang, C. Zhang, 
K. Wu, S. X. Dou, L. Chen, Z. Hu, and Y. Du,
Realization of flat band with possible nontrivial topology in electronic Kagome lattice,
Science Advances \textbf{4}, eaau4511 (2018).

\bibitem{leung20}
T.-H. Leung, M. N. Schwarz, S.-W. Chang, C. D. Brown, G. Unnikrishnan, and D. Stamper-Kurn,
Interaction-Enhanced Group Velocity of Bosons in the Flat Band of an Optical Kagome Lattice,
Phys. Rev. Lett. \textbf{125}, 133001 (2020).


\bibitem{diebel16}
F. Diebel, D. Leykam, S. Kroesen, C. Denz, and A. S. Desyatnikov,
Conical Diffraction and Composite Lieb Bosons in Photonic Lattices,
Phys. Rev. Lett. \textbf{116}, 183902 (2016).

\bibitem{kajiwara16}
S. Kajiwara, Y. Urade, Y. Nakata, T. Nakanishi, and M. Kitano,
Observation of a nonradiative flat band for spoof surface plasmons in a metallic Lieb lattice,
Phys. Rev. B \textbf{93}, 075126 (2016).

\bibitem{ozawa17}
H. Ozawa, S. Taie, T. Ichinose, and Y. Takahashi,
Interaction-Driven Shift and Distortion of a Flat Band in an Optical Lieb Lattice,
Phys. Rev. Lett. \textbf{118}, 175301 (2017).
%

\bibitem{barreteau17}
C. Barreteau, F. Ducastelle, and T. Mallah,
A bird's eye view on the flat and conic band world of the honeycomb and Kagome lattices: 
towards an understanding of 2D metal-organic frameworks electronic structure,
J. Phys.: Condens. Matter \textbf{29}, 465302 (2017).

\bibitem{mizoguchi19}
T. Mizoguchi and M. Udagawa,
Flat-band engineering in tight-binding models: Beyond the nearest-neighbor hopping,
Phys. Rev. B \textbf{99}, 235118 (2019).

\bibitem{twobandnote}
In the case of two-band lattices there are two distinct branches in the two-body 
spectrum. Assuming onsite attraction and uniform pairing, it can be shown that 
the lower (upper) branch is associated with the Goldstone (Leggett) modes 
that describe the perfectly in-phase (out-of-phase) collective fluctuations of the 
superfluid order parameter on two different sublattices~\cite{iskin21}. 
For this reason $\beta_{1\mathbf{q}} \propto (1,1)$ is a manifestation of the 
perfectly in-phase sublattice contribution to the pair formation.

\bibitem{topp21}
G. E. Topp, C. J. Eckhardt, D. M. Kennes, M. A. Sentef, and P. T\"orm\"a,
Light-matter coupling and quantum geometry in moir\'e materials,
Phys. Rev. B \textbf{104}, 064306 (2021).

\end{thebibliography}
\end{document}